\documentclass[aps,preprint,showpacs,preprintnumbers,amsmath,amssymb]{revtex4}

\usepackage{dcolumn}
\usepackage{mathrsfs}
\usepackage{bm}
\usepackage{amsmath,amssymb,epsfig,float}

\begin{document}

\title{Classical correlation and quantum discord sharing
of Dirac  fields in noninertial frame}

\author{Jieci Wang, Junfeng Deng and Jiliang Jing\footnote{Corresponding author, Email: jljing@hunnu.edu.cn}}
\affiliation{ Institute of Physics and  Department of Physics,
\\ Hunan Normal University, Changsha, \\ Hunan 410081, P. R. China
\\ and
\\ Key Laboratory of Low-dimensional Quantum Structures
\\ and Quantum
Control of Ministry of Education, \\ Hunan Normal University,
Changsha, Hunan 410081, P. R. China}

\vspace*{0.2cm}
\begin{abstract}
\vspace*{0.2cm} The classical and quantum correlations sharing
between modes of the Dirac fields in the noninertial frame are
investigated. It is shown that: (i)  The classical correlation for
the Dirac fields decreases as the acceleration increases, which is
different from the result of the scalar field that the classical
correlation is independent of the acceleration; (ii) There is no
simple dominating relation between the quantum correlation and
entanglement for the Dirac fields, which is unlike the scalar case
where the quantum correlation is always over and above the
entanglement;  (iii) As the acceleration increases, the correlations
between modes $I$ and $II$ and between modes $A$ and $II$ increase,
but the correlations between modes $A$ and $I$ decrease.
\end{abstract}

\vspace*{1.5cm}

\maketitle

\section{introduction}

The integration of the quantum information and another fundamental
part of modern physics theories---relativity theory gives birth to
the theory of the relativistic quantum information
\cite{Peres,Boschi,Bouwmeester}. It is believed that the
investigation of the quantum correlation in a relativistic framework
is not only helpful to understand some key questions in the quantum
information theory, but also plays an important role in the study of
the entropy and  information paradox \cite{Bombelli-Callen,
Hawking-Terashima} of the black hole. Following the pioneering work
presented by Peres {\it et al.}~\cite{SRQIT1}, many authors
\cite{Ging,Alsing-Mann, Schuller-Mann,Alsing-Milburn, Lamata,
Qiyuan,moradi,jieci1,Jason,Ralph,David,Andre,Eduardo1} considered
the quantum entanglement in a relativistic setting. Recently, Adesso
\emph{et al.} \cite{adesso} discussed the entanglement sharing of a
scalar field in the noninertial frame and found that the classical
correlation is independent of the acceleration of the observer if
the other observer stays stationary. Juan Le¨®n \emph{et al.}
studied the Dirac entanglement in both the spin and occupation
number cases \cite{Juan}, and Mann \emph{et al.} discussed the
speeding up entanglement degradation of the scalar field in the
noninertial frame \cite{Mann2}. Eduardo \emph{et al.} investigated
the the effect of the statistics on the entanglement \cite{Eduardo}
and showed that the entanglement survival of the Dirac field is
fundamentally inherent in the Fermi-Dirac statistics and that it is
independent of the number of the modes considered.

However, the entanglement is not the only characterization of a
quantum system, and it was found that it has no advantage in some
quantum information tasks. As seen in Refs.
\cite{datta-onequbit,experimental-onequbit}, although there is no
entanglement, certain quantum information processing tasks can also
be done efficiently. Such a resource for the quantum computation and
communication, i.e., the quantum correlation \cite{zurek,
vedral,zurek1}, is believed more practical than the entanglement.
Moreover, it could be used to improve the efficiency of the quantum
Carnot engine \cite{Carnot} and to get a better understand of the
quantum phase transition \cite{QPT}. More recently, Datta
\cite{datta} calculated the quantum correlation between two
relatively accelerated scalar modes, and showed that the quantum
correlation which is measured by the quantum discord, is over and
above the entanglement. In the limit of the infinite acceleration
there is a finite amount of quantum correlation while the
entanglement don't exist.

We noticed that most studies in the noninertial system focused on
the entanglement, while the study of the classical correlation and
how to distinguish the classical and quantum correlations is almost
ignored. In fact, this is an important problem for mixed states
since sometimes the quantum correlation is hidden by their classical
correlation \cite{QPT}. In this paper we will discuss the sharing of
the classical and quantum correlations of the Dirac fields in the
noninertial frame, as well as a comparative study of the
relationships between the entanglement and quantum correlation in
this system. We are interested in how the acceleration will
influence the classical and quantum correlations, and whether or not
the differences between Fermi-Dirac and Bose-Einstein statistic will
plays a role in the  classical and quantum correlations sharing. We
assume that two observers, Alice and Rob, share a entangled initial
state at the same point in flat Minkowski spacetime. After the
coincidence of them, Alice stays stationary while Rob moves with
uniform acceleration $a$. It is well known that a uniformly
accelerated observer in Rindler region $I$ has no access to the
field modes of the causally disconnected Rindler region $II$. Thus
we must trace over the inaccessible modes, which leads the initial
pure state to a mixed state. At the same time, from an inertial
perspective the system is bipartite, but from a noninertial
perspective an extra set of complementary modes in Rindler region
$II$ becomes relevant. Therefore, we have to calculate the classical
and quantum correlations in all possible bipartite divisions of the
tripartite system: the mode $A$ described by Alice, the mode $I$ in
Rindler region $I$ (described by Rob), and the complementary mode
$II$ in the Rindler region $II$.

The outline of the paper is as follows. In Sec. II we recall some
concept from the view of the quantum information theory, in
particular the classical and quantum correlations. In Sec. III we
investigate the essential features of the Dirac fields in the
noninertial frame. In Sec. IV we study the classical and quantum
correlations sharing in this system. We summarize and discuss our
conclusions in the last section.

\section{Classical and quantum correlations}

We now present a brief review of the classical correlation and quantum discord.
In classical information theory, the information can be quantified by Shannon entropy $H(X)=-\sum_x P_{|X=x} \log
P_{|X=x}$, where $P_{|X=x}$ is the probability with $X$ being $x$.
For two random variables $X$ and $Y$, the total correlation between them can be measured
by the mutual information which is defined as $I(X:Y) = H(X) + H(Y) -
H(X,Y)$, whose quantum version can be written as \cite{RAM}
\begin{equation}
{\cal I}(A:B)=S(\rho_A)+S(\rho_B)-S(\rho_{AB}), \label{mi2}
\end{equation}
where $S(\rho)=-{\rm Tr}(\rho {\rm log}\rho)$ is the von Neumann
entropy. By introducing the conditional entropy $H(Y|X)=H(Y,X)-H(X)$,
we can rewrite the mutual information as
\begin{equation}
I(X:Y)= H(Y) -  H(Y|X) \label{cmi3}.
\end{equation}
In order to generalize the above equation to the quantum domain, we
measure the subsystem $A$ of $\rho_{AB}$ by a complete set of
projectors $\{{\Pi_j}\}$, corresponds to the outcome $j$, which
yields $\rho_{B|j} ={Tr_A(\Pi_j\rho_{AB}\Pi_j)}/{p_j}$, with $
p_j=Tr_{AB}(\Pi_j\rho_{AB}\Pi_j)$. Then the quantum mutual
information can alternatively defined by
\begin{equation}
{\cal J}_{\{\Pi_j\}}(A:B)= S(\rho_B) -S_{\{\Pi_j\}}(B|A),
 \label{definition2}
\end{equation}
where $S_{\{\Pi_j\}}(B|A) = \sum_j p_j S(\rho_{B|j})$  is conditional
entropy \cite{Cerf} of the quantum state. The above quantity strongly depends on the choice of the
measurements $\{\Pi_j\}$. In order to calculate the classical correlation, we shall
minimize the conditional entropy over all possible measurements on $A$
which corresponds to finding the measurement that disturbs least
the overall quantum state and allows one to extract the most
information about the state  \cite{zurek}.
Then we define the classical correlation between parts
$A$ and $B$ as
\begin{equation}
{\cal C}(A:B)=\max_{\{\Pi_j\}}{\cal J}_{\{\Pi_j\}}(A:B),
\end{equation}
and the quantum discord \cite{zurek} as
\begin{equation}
{\cal D}(A:B)={\cal I}(A:B)-{\cal C}(A:B), \label{definition}
\end{equation}
which presents even for separable states and arises as a consequence
of coherence in a quantum system. In particular, a zero discord
means the information of the quantum state can be obtained by
observers without perturbing it, i.e, all the correlations are
classical.

\section{Quantization for Dirac filed in Minkowski spacetime and
accelerated system}

For an inertial observer in flat Minkowski spacetime,
the field can be quantized in a straightforward manner by expanding it
in terms of a complete set of positive and negative frequency modes
\begin{eqnarray}
\Psi = \int d k \, ( a_{\mathbf{k}} \, \psi^+_\mathbf{k} + b^{\dag}_{\mathbf{k}} \, \psi^-_\mathbf{k} ),
\end{eqnarray}
where $\mathbf{k}$ is the wave vector which is used to label the
modes and for massless Dirac fields $\omega=|\mathbf{k}|$.
The above positive and negative frequency modes satisfy $\{
a_{\mathbf{k}}, a^{\dag}_{\mathbf{k}'}\}=\{
b_{\mathbf{k}},b^{\dag}_{\mathbf{k}'}\}=\delta(\mathbf{k}-\mathbf{k}')$
with all other anticommutators vanishing.

The appropriate coordinates to describe Rob's motion are the Rindler
coordinates, given by
\begin{eqnarray}
at=e^{a \varepsilon}\ \sinh {(a\eta)}, \ \ \ \   az=e^{a \varepsilon}\ \cosh {(a\eta)}.
\end{eqnarray}
The quantum field theory for the Rindler observer is constructed by expanding
the field in terms of the complete set of positive and negative frequency modes \cite{Alsing-Mann}
\begin{eqnarray}\label{First expand}
&&\Psi=\int
d\mathbf{k}[\hat{c}^{I}_{\mathbf{k}}\Psi^{I+}_{\mathbf{k}}+\hat{d}^{I\dag}_{\mathbf{k}}\Psi^{I-}
_{\mathbf{k}}+\hat{c}^{II}_{\mathbf{k}}\Psi^{II+}_{\mathbf{k}}+\hat{d}^{II\dag}_{\mathbf{k}}\Psi^{II-}_{\mathbf{k}}],
\end{eqnarray}
where $\hat{c}^{I}_{\mathbf{k}}$ and $\hat{d}^{I\dag}_{\mathbf{k}}$
are the fermion annihilation and antifermion creation operators
acting on the state in region $I$, and
$\hat{c}^{II}_{\mathbf{k}}$ and $\hat{d}^{II\dag}_{\mathbf{k}}$ are
the fermion annihilation and antifermion creation operators acting
on the state in region $II$ respectively. The canonical anticommutation relations of the mode operators are
\begin{eqnarray}
&&\{\hat{c}^{I}_{\mathbf{k}},\hat{c}^{I\dag}_{\mathbf{k'}} \}=\{\hat{d}^{I}_{\mathbf{k}},\hat{d}^{I\dag}_{\mathbf{k'}} \}
=\delta(\mathbf{k}-\mathbf{k'}),\\ \nonumber
&&\{\hat{c}^{II}_{\mathbf{k}},\hat{c}^{II\dag}_{\mathbf{k'}} \}=\{\hat{d}^{II}_{\mathbf{k}},\hat{d}^{II\dag}_{\mathbf{k'}} \}
=\delta(\mathbf{k}-\mathbf{k'}),
\end{eqnarray}
with all other anticommutators vanishing. All the above positive and
negative frequency modes are defined using the future-directed
timelike Killing vector: in inertial frame the Killing vector is
$\partial_t$, in Rindler region $I$ is $\partial_\eta$ and in region
$II$ is $\partial_{-\eta}$.

We can easily get the Bogoliubov transformations \cite{Barnett} between the
creation and annihilation operators of Rindler and Minkowski
coordinates. After properly normalizing the state vector, the
Minkowski vacuum is found to be an entangled two-mode squeezed state
\begin{eqnarray}\label{Dirac-vacuum}
|0_{\mathbf{k}}\rangle_{M}= \cos
r|0_{\mathbf{k}}\rangle_{I}|0_{-\mathbf{k}}\rangle _{II}+\sin
r|1_{\mathbf{k}}\rangle_{I}|1_{-\mathbf{k}}\rangle _{II},
\end{eqnarray}
where $\cos r=(e^{-2\pi\omega/a}+1)^{-\frac{1}{2}}$ with $a$ is Rob's
acceleration.
We noticed that in scalar case the Minkowski vacuum
is $|0_{\mathbf{k}}\rangle_{M}= \frac{1}{\cosh r}
\sum_{n=0}^{\infty}\tanh^n r|n_{\mathbf{k}}\rangle_{I}|n_{-\mathbf{k}}
\rangle _{II}$ with $\cosh r=(1-e^{-2\pi\omega/a})^{-\frac{1}{2}}$.
The disparity between the scalar field and Dirac field is caused by the differences between Fermi-Dirac
and Bose-Einstein statistic. Due to the Pauli exclusion principle, there are only two allowed
states for each mode, $|0\rangle_{M}$ and $|1\rangle_{M}$
for fermions, and similarly for antifermions. Thus, the only
excited state is given by
\begin{eqnarray}\label{Dirac-excited}
|1\rangle_{M}=|1_{\mathbf{k}}\rangle_{I}|0_{-\mathbf{k}}\rangle_{II}.
\end{eqnarray}
For simplicity we refer to the particle mode
$\{|n_{\mathbf{k}}\rangle_{I}\}$ simply as $\{|n\rangle_{I}\}$, and
the anti-particle mode $\{|n_{-\mathbf{k}}\rangle_{II}\}$ as
$\{|n\rangle_{II}\}$.
When Rob travels with uniform acceleration through
the Minkowski vacuum, his detector registers the number of particles
\begin{eqnarray}
N^2=\frac{1}{e^{2\pi\omega/a}+1},
\end{eqnarray}
which shows the accelerated observer detects a thermal
Fermi-Dirac distribution of particles.

\section{classical and quantum correlations sharing}

The redistribution of entanglement in tripartite systems has been studied both
in inertial \cite{Coffman} and noninertial systems \cite{Alsing-Mann,adesso} by
tracing over any one of the three qubits to calculate all the bipartite entanglement.
Here we will use a similar method to discuss the quantum  and
classical correlation sharing in the accelerated fermi system.
We assume that Alice has a detector which only detects mode
$|n\rangle_{A}$ and Rob has a detector sensitive only to mode
$|n\rangle_{R}$, they share a maximally entangled initial state
\begin{eqnarray}\label{initial}
|\Phi\rangle_{AR}=\frac{1}{\sqrt{2}}(|0\rangle_{A}|0\rangle_{R}
+|1\rangle_{A}|1\rangle_{R}),
\end{eqnarray} at the
same point in flat Minkowski spacetime.
After the coincidence of Alice and Rob, Alice stays
stationary while Rob moves with uniform acceleration $a$. Using Eqs.
(\ref{Dirac-vacuum}) and (\ref{Dirac-excited}), we can rewrite Eq.
(\ref{initial}) in terms of Minkowski modes for Alice and Rindler
modes for Rob
\begin{eqnarray}  \label{state}
\nonumber|\Phi\rangle_{A,I,II}&=&\frac{1}{\sqrt{2}}( \cos r|0\rangle_{A}
|0\rangle_{I}|0\rangle_{II}+\sin r|0\rangle_{A}
|1\rangle_{I}|1\rangle_{II}\\&&+|1\rangle_{A}
|1\rangle_{I}|0\rangle_{II}).
\end{eqnarray}

\subsection{Physical accessible correlations}

Since Rob is causally disconnected from the region $II$, the only
information which is physically accessible to the observers is
encoded in the mode $A$ described by Alice and the mode $I$
described by Rob. Taking the trace over the state of region $II$, we
obtain
\begin{eqnarray}  \label{eq:state1}
\nonumber\rho_{A,I}=\frac{1}{2}\bigg[\cos^2
r|00\rangle\langle00|+\cos
r(|00\rangle\langle11|+|11\rangle\langle00|) +\sin^2
r|01\rangle\langle01|+|11\rangle\langle11|\bigg],
\end{eqnarray}
where $|mn\rangle=|m\rangle_{A}|n\rangle_{I}$. The von Neumann
entropy of this state is $S(\rho_{A,I})=-\frac{1+\cos^2
r}{2}\log_2(\frac{1+\cos^2 r}{2})-\frac{1-\cos^2
r}{2}\log_2(\frac{1-\cos^2 r}{2})$. Similarly, we can obtain the
entropy $S(\rho_A)$ for the reduced density matrix of the mode $A$
and $S(\rho_{II})$ for the mode $II$, respectively.

Then let us make our measurements on the $A$ subsystem, the projectors
are defined as \cite{zurek,datta-onequbit}
\begin{eqnarray}
\Pi_{+}=\frac{I_2+\mathbf{n}\cdot\sigma}{2}\otimes I_2,\ \ \ \ \
\Pi_{-}=\frac{I_2-\mathbf{n}\cdot\sigma}{2}\otimes I_2,
\end{eqnarray}
where $n_1=\sin\theta\cos\varphi,\ n_2=\sin\theta\sin\varphi,\
n_3=\cos\theta$ and $\sigma_i$ are Pauli matrices. After the
measurement of $\Pi_{+}$, the quantum state $\rho_{A,I}$ changes to
\begin{eqnarray}\label{con1}
\nonumber\rho_{(I|+)}&=&Tr_{A}(\Pi_{+}\rho_{A,I}\Pi_{+})/p_{+}=Tr_{A}(\Pi_{+}\rho_{A,I})/p_{+}\\
\nonumber&=&                              \frac{1}{4p_{+}}Tr_A\left[\begin{array}{cccc}
                                                     (1+\cos\theta)\cos^2 r  & e^{i\varphi}\sin\theta \cos r & e^{-i\varphi}\sin\theta \cos^2 r & (1-\cos\theta)\cos r  \\
                                                     0 & (1+\cos\theta)\sin^2 r & 0 & e^{-i\varphi}\sin\theta \sin^2 r \\
                                                     0 & 0 & 0 & 0 \\
                                                     (1+\cos\theta)\cos r & e^{i\varphi}\sin\theta  & e^{-i\varphi}\sin\theta \cos r & 1-\cos\theta \\
                                                   \end{array}
                                                 \right]\\
&=&\frac{1}{2}\left[
     \begin{array}{cc}
       (1+\cos\theta)\cos^2 r & e^{i\varphi} \cos r\sin\theta  \\
       e^{-i\varphi} \cos r\sin\theta  & 1-\cos\theta+(1+\cos\theta)\sin^2 r \\
     \end{array}
   \right],
\end{eqnarray}
where $p_{+}=Tr(\Pi_{+}\rho_{A,I} \Pi_{+})=1/2$, $\Pi_{+}^2=\Pi_{+}$, and
$Tr(\alpha\beta)=Tr(\beta\alpha)$ were used. The eigenvalues of this density matrix are
$\lambda_+(1,2)=\frac{1}{2}(1\pm\sqrt{1\pm\sin^2 2r\cos^4 \frac{\theta}{2}})$.
Similarly, we can obtain
\begin{eqnarray}\label{con2}
\rho_{(I|-)}=\frac{1}{4p_{-}}\left[
     \begin{array}{cc}
       (1-\cos\theta)\cos^2 r & -e^{i\varphi} \cos r \sin\theta\\
       -e^{-i\varphi} \cos r\sin\theta & 1+\cos\theta+(1-\cos\theta)\sin^2 r\\
     \end{array}
   \right],
\end{eqnarray}
where $p_{-}=Tr(\Pi_{-}\rho_{A,I} \Pi_{-})=1/2$, which yields to
$\lambda_-(1,2) =\frac{1}{2}(1\pm\sqrt{1\pm\sin^2 2r\sin^4
\frac{\theta}{2}})$. Using Eqs. (\ref{con1}) and (\ref{con2}), we
obtain the conditional entropy $S_{\{\Pi_j\}}(I|A)\equiv\sum_jp_j
S(I|j)$. The classical correlation  in this case is
\begin{eqnarray}\label{classical1}
{\cal C}(\rho_{A,I})=-\frac{\cos^2 r}{2}\log_2(\frac{\cos^2 r}{2})
-(1-\frac{\cos^2 r}{2})\log_2(1-\frac{\cos^2 r}{2})
-\min_{\Pi_j}S_{\{\Pi_j\}}(I|A),
\end{eqnarray}
and the value of quantum discord can be given by
\begin{eqnarray}\label{discord1}
\nonumber {\cal D}(\rho_{A,I})=1&+&\frac{1+\cos^2 r}{2}\log_2(\frac{1+\cos^2 r}{2})
+\frac{1-\cos^2 r}{2}\log_2(\frac{1-\cos^2 r}{2})\\
&+&\min_{\Pi_j}S_{\{\Pi_j\}}(I|A).
\end{eqnarray}

\begin{figure}[ht]
\includegraphics[scale=1.0]{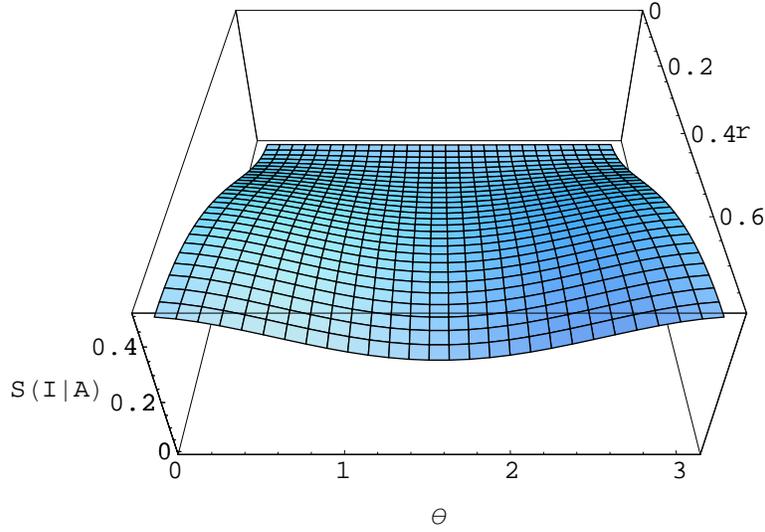}\vspace{0.0cm}
\caption{\label{Cla1}(Color online) The condition entropy $S(I|A)$
as functions of $r$ and $\theta$.}
\end{figure}

\begin{figure}[ht]
\includegraphics[scale=1.0]{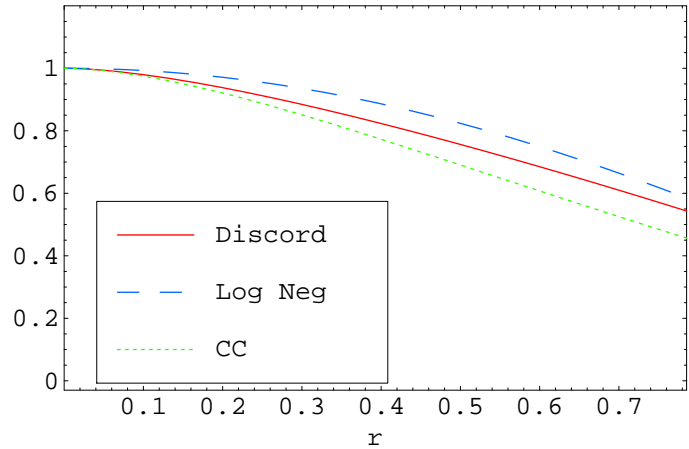}
\caption{\label{Dl1}(Color online) The discord(red line), logarithmic negativity (dashed blue line),
and classical correlation (dotted green line) of $\rho_{A,I}$
as a function of $r$.}
\end{figure}

Note that the conditional entropy has to be numerically evaluated by
optimizing over the angles $\theta$ and $\phi$. Thus we should
minimize it over all possible measurements on $A$, which corresponds
to finding the measurement that disturbs least the overall quantum
state. It is clear that the condition entropy is independent of
$\varphi$, thus we plot it as functions of $r$ and $\theta$ in Fig.
\ref{Cla1}, from which the minimum of the conditional entropy can be
obtained when $\theta=\pi/2$. Then we can get the accurate value of
Eqs. (\ref{classical1}) and (\ref{discord1}) and plot them in Fig.
\ref{Dl1}. For comparison, we also plot the logarithmic negativity
\cite{Vidal} of the same state.

Fig. \ref{Dl1} shows how the acceleration changes the classical and
quantum correlations. The monotonous decrease of ${\cal
D}(\rho_{A,I})$ as the acceleration increases means that the quantum
correlation of state $\rho_{A,I}$ decreases due to the thermal
fields generated by the Unruh effect. Note that  the classical
correlation for the Dirac fields ${\cal C}(\rho_{A,I})$ also
decreases as the acceleration increases, which is different from the
result of scalar field  that the classical correlation is
independent of the acceleration \cite{adesso}. It is interesting to
note that in the Dirac case entanglement of $\rho_{A,I}$ is always
larger than quantum correlation, which is in sharply contrast to the
result of the scalar fields that the quantum correlation is always
over and above entanglement \cite{datta}. This obvious distinction
is caused by the differences between Fermi-Dirac and Bose-Einstein
statistics. The Dirac particles which have half-integer spin must
obey the Pauli exclusion principle and access to only two quantum
levels.

\subsection{Physical unaccessible correlations}

To explore correlations in this system in detail we consider
the tripartite system consisting of the modes $A$, $I$, and $II$.
We therefore calculate the correlations in all possible bipartite
divisions of the system. Let us first comment on the
correlations created between the mode $A$ and mode $II$, tracing
over mode $I$ of the state Eq. (\ref{state}) , we obtain the density matrix
\begin{eqnarray}  \label{eq:state}
\nonumber\rho_{A,II}=\frac{1}{2}\bigg[\cos r^2|00\rangle\langle00|+
\sin r(|10\rangle\langle01|+|01\rangle\langle10|) +\sin^2
r|01\rangle\langle01|+|10\rangle\langle10|\bigg],
\end{eqnarray}
where $|mn\rangle=|m\rangle_{A}|n\rangle_{II}$. We can easily get
$S(\rho_{A,II})=-\frac{1}{2}\cos^2 r\log_2\frac{\cos^2
r}{2}-(1-\frac{1}{2} \cos^2 r)\log_2(1-\frac{1}{2}\cos^2 r)$. After
these measurements, the state $\rho_{A,II}$  changes to
\begin{eqnarray}\label{con3}
\frac{1}{4p'_{\pm}}\left[
     \begin{array}{cc}
       1\mp\cos\theta+(1\pm\cos\theta)\cos^2 r & \pm e^{i\varphi} \sin r \sin \theta \\
      \pm e^{-i\varphi} \sin r\sin \theta & (1\pm\cos\theta)\sin^2 r \\
     \end{array}
   \right],
\end{eqnarray}
where $p'_{+}=p'_{-}=1/2$.
According to the preceding calculations, we have
\begin{eqnarray}\label{classcal2}
{\cal C}(\rho_{A,II})=-\frac{1+\cos^2 r}{2}\log_2(\frac{1+\cos^2 r}{2})
-\frac{1-\cos^2 r}{2}\log_2(\frac{1-\cos^2 r}{2})
-\min_{\Pi_j}S_{\{\Pi_j\}}(II|A),
\end{eqnarray}
and
\begin{eqnarray}\label{discord2}
\nonumber {\cal D}(\rho_{A,II})=1&+&\frac{1}{2}\cos^2 r\log_2\frac{\cos^2 r}{2}
+(1-\frac{1}{2}\cos^2 r)\log_2(1-\frac{1}{2}\cos^2 r)\\
&+&\min_{\Pi_j}S_{\{\Pi_j\}}(II|A).
\end{eqnarray}
Similarly, the minimum of the conditional entropy
$S_{\{\Pi_j\}}(II|A)\equiv\sum_jp_j S(II|i)$
can be obtained when $\theta=\pi/2$.
Tracing over the modes in $A$, we obtain the density matrix
\begin{eqnarray}
\nonumber\rho_{I,II}=\frac{1}{2}\bigg[\cos r^2|00\rangle\langle00|+
\sin r \cos r(|00\rangle\langle11|+|11\rangle\langle00|)
+|10\rangle\langle10|+\sin^2 r|11\rangle\langle11|\bigg],
\end{eqnarray}
where $|mn\rangle=|m\rangle_{I}|n\rangle_{II}$.
The von Neumann entropy of this matrix is
$S(\rho_j)=-\frac{1}{2}\cos^2 r\log_2\frac{\cos^2 r}
{2}-(1-\frac{1}{2}\cos^2 r)\log_2(1-\frac{1}{2}\cos^2 r)$.
After those measurements, the state $\rho_{I,II}$  changes to
\begin{eqnarray}\label{con33}
\frac{1}{4p''_{\pm}}\left[
     \begin{array}{cc}
       1\mp\cos\theta+(1\pm\cos\theta)\cos^2 r & \pm e^{i\varphi}\sin r \cos r \sin \theta \\
       \pm e^{-i\varphi}\sin r \cos r\sin \theta & (1\mp\cos\theta)\sin^2 r \\
     \end{array}
   \right],
\end{eqnarray}
where $p''_{\pm}=\frac{1}{2}(1\mp\cos \theta \sin^2 r)$.
The classical correlation of $\rho_{I,II}$ is
\begin{eqnarray}\label{classcal3}
{\cal C}(\rho_{I,II})=-\frac{1+\cos^2 r}{2}\log_2(\frac{1+\cos^2
r}{2}) -\frac{1-\cos^2 r}{2}\log_2(\frac{1-\cos^2 r}{2})
-\min_{\Pi_j}S_{\{\Pi_j\}}(II|I),
\end{eqnarray}
and the quantum correlation is
\begin{eqnarray}\label{discord3}
\nonumber {\cal D}(\rho_{I,II})=&-&\frac{1}{2}\cos^2 r\log_2
\frac{\cos^2 r}{2}-(1-\frac{1}{2}\cos^2 r)\log_2(1-\frac{1}{2}\cos^2 r)-1\\
&+&\min_{\Pi_j}S_{\{\Pi_j\}}(II|I).
\end{eqnarray}
The only difference is that we can get the minimum of $S_{\{\Pi_j\}}(II|I)\equiv\sum_jp_j S(II|j)$
when $\theta=\pi/4$.

\begin{figure}[ht]
\includegraphics[scale=1.0]{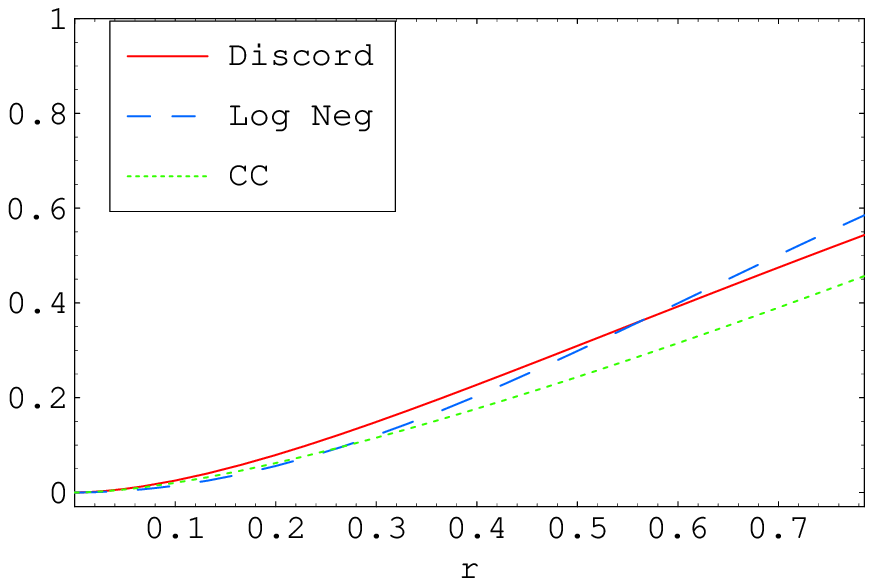}
\caption{\label{Dl2} (Color online) The discord(red line), logarithmic negativity(dashed blue line),
and classical correlation(dotted green line) of $\rho_{A,II}$
as a function of $r$.}
\end{figure}
\begin{figure}[ht]
\includegraphics[scale=1.0]{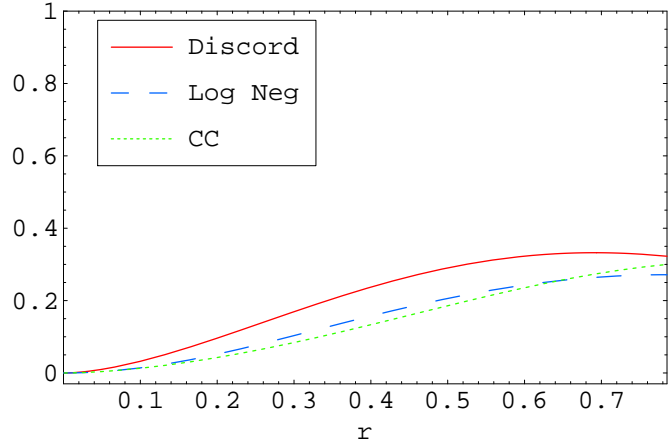}
\caption{\label{DL3}(Color online) The discord(red line), logarithmic negativity(dashed blue line),
and classical correlation(dotted green line) of $\rho_{I,II}$
as a function of $r$.}
\end{figure}

The properties of the correlations of $\rho_{A,II}$ and
$\rho_{I,II}$ are shown in Figs.(\ref{Dl2}) and (\ref{DL3}). They
demonstrate that both the classical and quantum correlations of
these two states increase as the acceleration increases. It is
interesting to note that, for the state $\rho_{A,II}$, the quantum
correlation dominates the entanglement when $r$ is small, and the
dominance is reversed as the acceleration increases. While for the
state $\rho_{I,II}$, the quantum correlation always dominates the
entanglement. That is to say, the quantum correlation can be larger
than the entanglement for some states, whereas it can be smaller for
other states. Thus, we arrive at the conclusion that there is no
simple dominating relation between the quantum correlation and the
entanglement in the noninertial frame.

\begin{figure}[ht]
\includegraphics[scale=1.0]{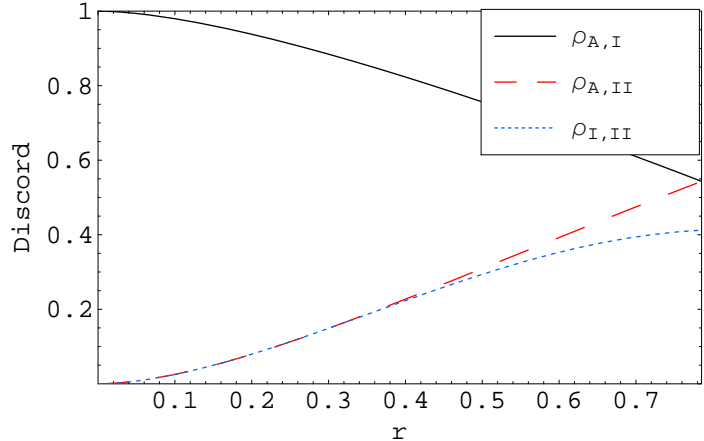}
\caption{\label{Dis}(Color online) The quantum discords $D(\rho_{a,b})$ of
$\rho_{A,I}$(black line), $\rho_{A,II}$(dashed red line) and $\rho_{I,II}$(dotted blue line) as a function
of $r$. In the limit of infinite acceleration, i.e. $r\rightarrow \pi/4$,
the discord of $\rho_{A,I}$ equals to the discord of $\rho_{A,II}$.}
\end{figure}

\begin{figure}[ht]
\includegraphics[scale=1.0]{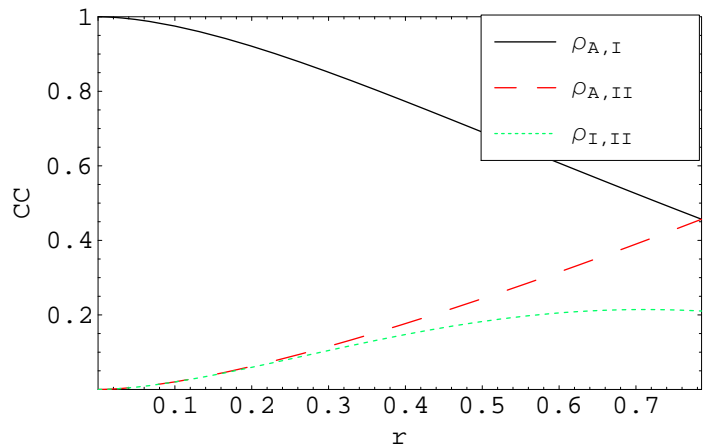}
\caption{\label{Class}(Color online) The classical correlations of
$\rho_{A,I}$(black line), $\rho_{A,II}$(dashed red line) and $\rho_{I,II}$ (dotted green line) as a function
of $r$. In the case of $r\rightarrow \pi/4$,
the ${\cal C}(\rho_{A,I})$ also equals to ${\cal C}(\rho_{A,II})$.}
\end{figure}

In Figs. (\ref{Dis}) and (\ref{Class})  we plot the redistributions
of the quantum and classical correlation which show how the
acceleration changes all the bipartite correlations. For lower
acceleration, modes $A$ and $I$ remain almost maximally correlated
while there is little correlations between modes $I$ and $II$ and
between modes $A$ and $II$. As the acceleration grows, the
unaccessible correlations between modes $I$ and $II$ and between
modes $A$ and $II$ increase, while the accessible correlation
between modes $A$ and $I$ decreases. Consequently, the original two
mode correlations in the state Eq. (\ref{initial}) described by
Alice and Rob from an inertial perspective, are redistributed among
the mode $A$ described by Alice, the mode $I$ described by Rob, and
the complementary mode $II$. That is to say, the initial
correlations described by inertial observers are redistributed
between all the bipartite modes. Therefore, as a consequence of the
monogamy of correlations, the physically accessible correlations
between the two modes described by Alice and Rob degrade. In the
limit of infinite acceleration, the correlations of $\rho_{A,I}$
equal to the correlations of $\rho_{A,II}$. It is worth to
mention that the authors in Ref. \cite{datta} only discussed the
quantum correlation between Alice and the Rindler region $I$, while
we discussed the redistribution of quantum correlation and classical
correlation among all, accessible and unaccessible modes. The
analysis of the unaccessible correlations between Alice and
region $II$, as well as regions $I$ and $II$ were used to
explain the loss of accessible correlations between Alice
and region $I$. The loss of correlations in accessible modes
were redistributed to the unaccessible modes.

\section{summary}

The effect of the acceleration on the redistribution of the
classical and quantum correlations  of the Dirac fields in the
noninertial frame is investigated. It is shown that: (i) The
classical correlation ${\cal C}(\rho_{A,I})$ for the Dirac fields
decreases as the acceleration increases, which is different from the
result of the scalar case that the classical correlation is
independent of the acceleration \cite{adesso}. (ii) In the Dirac
case, we find that the entanglement always dominates the quantum
correlation for the state $\rho_{A,I}$, the dominating relation of
the entanglement and quantum correlation is reversed at a point as
the acceleration increases for the state $\rho_{A,II}$, while the
quantum correlation always dominates the entanglement for the state
$\rho_{I,II}$. These results are in sharply contrast to the result
of the scalar case that the quantum correlation is always over and
above the entanglement \cite{datta}. Thus, there is no simple
dominating relation between the quantum correlation and the
entanglement for the Dirac fields in the noninertial system. And
(iii) as the acceleration increases, the correlations between modes
$I$ and $II$ and between modes $A$ and $II$ increase, while the
correlations between modes $A$ and $I$ decrease. Thus, the original
correlations described by Alice and Rob from an inertial perspective
are redistributed between all the bipartite modes from a noninertial
perspective.

If we consider Alice to be accelerated as well, the increase
and decrease of the correlations would be more obvious
as the acceleration increases. The results of this paper
can be also applied to the case that Alice stays stationary at the
asymptotically flat region of a black hole while Rob barely escapes through it with
eternal uniform acceleration. Such topics are left for a future
research.

\begin{acknowledgments}
This work was supported by the National Natural Science Foundation
of China under Grant No. 10875040,  the key project of the National
Natural Science Foundation of China under Grant No 10935013, the
National Basic Research of China under Grant No. 2010CB833004,  the
Hunan Provincial Natural Science Foundation of China under Grant No.
08JJ3010, PCSIRT, No. IRT0964, and Construct Program of the National
Key Discipline.
\end{acknowledgments}

\end{document}